\documentstyle[12pt]{article}
\newcommand{\nc}{\newcommand}
\nc{\drm}{\rm d}
\nc{\pp}{/\!\!\!p} \nc{\bt}{\beta} \nc{\bb}{\begin{equation}} \nc{\ee}{\end{equation}}
\nc{\vecna}{\mbox{\boldmath $\nabla$}} \nc{\ga}{\gamma} \nc{\ia}{{\bf i}}
\nc{\age}{\dagger} \nc{\sig}{\sigma} \nc{\var}{\varphi} \nc{\longr}{\longrightarrow}
\nc{\al}{\alpha} \nc{\vare}{\varepsilon} \nc{\C}{C\!\!\!\!C} \nc{\1}{1\!\!1}
\nc{\lan}{\langle} \nc{\ran}{\rangle} \nc{\R}{I\!\!R} \nc{\h}{\hspace*{0.5 cm}}
\nc{\hs}{\hspace*} \nc{\wide}{\widehat} \nc{\pa}{\partial} \nc{\cent}{\centerline}
\nc{\vs}{\vspace*} \nc{\Vbf}{\mbox{\boldmath $V$}} \nc{\Abf}{\mbox{\boldmath $A$}} \nc{\sbf}{\mbox{\boldmath $s$}}
\nc{\jbf}{\mbox{\boldmath $j$}} \nc{\pabf}{\mbox{\boldmath $\partial$}}
\nc{\albf}{\mbox{\boldmath $\alpha$}} \nc{\sigbf}{\mbox{\boldmath $\sigma$}}
\nc{\gabf}{\mbox{\boldmath $\gamma$}} \nc{\ombf}{\mbox{\boldmath $\omega$}}
\nc{\vbf}{\mbox{\boldmath $v$}} \nc{\abf}{\mbox{\boldmath $a$}}
\nc{\bbf}{\mbox{\boldmath $b$}} \nc{\cbf}{\mbox{\boldmath $c$}}
\nc{\wbf}{\mbox{\boldmath $w$}} \nc{\xbf}{\mbox{\boldmath $x$}}
\nc{\xibf}{\mbox{\boldmath $\xi$}} \nc{\Xbf}{\mbox{\boldmath $X$}}
\nc{\ubf}{\mbox{\boldmath $u$}} \nc{\rbf}{\mbox{\boldmath $r$}}
\nc{\imp}{\mbox{\boldmath $p$}} \nc{\ug}{\; = \;}
\nc{\psib}{\overline{\psi}} \nc{\psip}{\psi^{\dagger}}
\nc{\psis}{\psi^{\star}} \nc{\G}{{\wide G}} \nc{\X}{{\wide X}} \nc{\A}{{\wide A}}
\nc{\Ho}{{\wide H}} \nc{\vo}{{\wide {\vbf}}} \nc{\jo}{{\wide {\jbf}}}
\nc{\xo}{{\wide {\xbf}}} \nc{\po}{{\wide {p}}} \nc{\impo}{\wide {\imp}}
\nc{\xio}{{\wide {\xibf}}} \nc{\Xo}{{\wide {\Xbf}}} \nc{\eto}{{\wide {\eta}}}
\nc{\xdo}{{\wide {{\dot {\xbf}}}}} \nc{\wo}{{\wide {\wbf}}}
\nc{\ao}{{\wide {\abf}}} \nc{\ro}{{\wide {\rbf}}} \nc{\Vo}{{\wide {\Vbf}}}
\nc{\xido}{{\wide {{\dot {\xibf}}}}} \nc{\Xdo}{{\wide {{\dot {\Xbf}}}}}
\nc{\hh}{{\hbar\over 2}} \nc{\bi}{\bibitem} \nc{\erm}{{\rm {e}}}

\begin{document}

\baselineskip 0.7cm

\cent{\bf VELOCITY FIELD AND OPERATOR FOR SPINNING PARTICLES}

\vs{0.3 cm}

\cent{\bf IN (NON-RELATIVISTIC) QUANTUM MECHANICS$^{\dagger}$}

\footnotetext{$^{\dagger}$ Work partially supported by INFN, CNR, MURST,
and by CNPq.}

\vs{0.1cm}

\begin{center}

{Giovanni SALESI

{\em Dipartimento di Fisica, Universit\`a  Statale di Catania,
95129--Catania, Italy; \ and

INFN--Sezione di Catania, Catania, Italy.}

\vs{0.2 cm}

and

\vs{0.2 cm}

Erasmo RECAMI

{\em Facolt\`a di Ingegneria, Universit\`a Statale di Bergamo,
24044--Dalmine (BG), Italy;

INFN--Sezione di Milano, Milan, Italy; \ and

Dept. of Applied Math., State University at Campinas, Campinas,
Brazil.}}

\end{center}

\vs{1.4cm}

\baselineskip 0.5cm

{\bf ABSTRACT \ ---} \ Starting from the formal expressions of the
hydrodynamical (or
``local'') quantities employed in the applications
of Clifford Algebras to quantum mechanics, we introduce ---{\em in terms of
the ordinary
tensorial framework}--- a new definition for the {\em field} of a generic
quantity. \ By translating from Clifford into tensor algebra, we also propose a new
(non-relativistic) velocity operator for a spin ${1\over 2}$ particle. This
operator is the sum of the ordinary
part $\imp/m$ describing the {\em mean} motion (the motion of the
center-of-mass), and of a second part associated with the so-called
{\em zitterbewegung}, which is the spin ``internal'' motion observed in the
center-of-mass frame. This spin component of the velocity operator is non-zero
not only in the {\em Pauli} theoretical framework, i.e. in presence of external magnetic fields
and spin precession, but also in the {\em Schr\"odinger} case, when the
wave-function is a spin eigenstate. In the latter case, one
gets a decomposition of the velocity field for the {\em Madelung
fluid} into two distinct parts: which constitutes
the non-relativistic analogue of the {\em Gordon decomposition}
for the Dirac current. \ We find furthermore that the zitterbewegung motion involves a
velocity field which is solenoidal, and that the local angular
velocity is parallel to the spin vector. In presence of a non-constant spin
vector (Pauli case) we have, besides the component normal to spin present even
in the Schr\"odinger theory, also a component of the local velocity which is
parallel to the rotor of the spin vector.

\vfill

\newpage

\baselineskip 0.9cm

{\bf 1. -- Hydrodynamical observables in quantum theory}\\

\h The {\em Multivector} or {\em Geometric} Algebras are essentially
due to
the work of great mathematicians of the nineteenth century as Hamilton
(1805--1863), Grassmann (1809--1877) and mainly Clifford (1845--1879).
  \ More recently, starting
from the sixties, some
authors, and in particular Hestenes,$^{[1-3]}$ did sistematically
study various interesting physical applications of such algebras, and
especially of the Real Dirac Algebra $\R_{1,3}$, often renamed
{\em Space-Time Algebra} (STA).$^{[4-6]}$ \ Rather interesting appear,
in microphysics, the applications to space-time [O(3), Lorentz]
transformations, to gauge [SU(2), SU(5), strong and electroweak
isospin] transformations, to chiral [SU(2)$_L$] transformations, to the
Maxwell equations, magnetic monopoles,$^{[7]}$ and so on.  \
But the most rich and rigorous application is probably the formal and
conceptual
analysis of the geometrical, kinematical and hydrodynamical content of
the Pauli and Dirac equations, performed by means
of the Real Pauli and Real Dirac Algebras, respectively. \  We shall refer
ourselves to non-relativistic physics, and therefore adopt
the Real Pauli Algebra, which is known to be {\em isomorphic to the ordinary tensorial
algebra of the so-called Pauli matrices} [SU(2)]. \ In this paper, when
ambiguities arise, the operators
will be distinguished by a cap.

\h As well-known, in the usual {\em hydrodynamical picture of fluids},
every physical quantity
depends not only on time, but also on the considered space point.
In other words, every quantity is a {\em local} or {\em field}
quantity:
\bb
G \equiv G(x) \qquad x = (t;{\bf x}) \; .
\ee
In the Pauli Algebra the {\em local} value of $G$ may be expressed as follows:
\bb
G(x) \ug \frac{\langle \psi\G\psip \rangle_0}{\psi\psip}\; ,
\ee
where $\langle \ \rangle_0$
indicates the {\em scalar part} of the {\em Clifford product} of the quantities
appearing within the brackets. \ \
Let us translate eq.(2) into the ordinary tensorial language:
\bb
G(x) \; = \; \frac{{\rm Re}\,[\psip\G\psi]}{\psip\psi}\; .
\ee
It is easy to see, and remarkable, that this operator definition for
$G(x)$ is equivalent
to the {\em real part} of the so-called dual representation for
{\em bilinear} operators, sometimes utilized in the literature.$^{[8-10]}$ \
In the ordinary approaches, such operators are commonly, even if
{\em implicitly} only, employed
for obtaining the probability densities of various quantities entering
the Schr\"odinger, Klein--Gordon or Dirac wave-equations. In this sense, one
can say that definition (3) agrees with the theoretical
apparatus of ordinary wave-mechanics.

\h In connection with the first two mentioned wave-equations (and in the
Dirac case, when confining ourselves to the translational--convective
part of the well-known Gordon decomposition$^{[11]}$),
the energy density may be put into the following form:
\bb
\frac{i\hbar}{2}\,[\psip(\pa_t\psi) - (\pa_t\psip)\psi]\; \equiv \;
\frac{1}{2}\psip i\hbar\stackrel{\leftrightarrow}{\partial_t} \psi\; ,
\ee
as easily obtained from eq.(3) for the hamiltonian \ $\G \equiv H \equiv
i\hbar\pa_t$. \ \
Analogously, for the current density one can write:
\bb
- \frac{i\hbar}{2m}\,[\psip(\vecna\psi) - (\vecna\psip)\psi]\; \equiv \;
\frac{1}{2m}\psip (-i\hbar\stackrel{\leftrightarrow}{\vecna}) \psi\; ,
\ee
as required by eq.(3) if \ $G \equiv \imp/m ;\, \G \equiv - i\hbar\vecna/m$. \ \
Therefore, the use of the bilinear operators \ $\stackrel{\leftrightarrow}
{\partial_t}$ \
and \ $\stackrel{\leftrightarrow}{\vecna}$ \ does allow us to write the
above densities in the form expected for quantum-mechanical densities:
namely in the form $\psip\X\psi$.

\h Let us notice that, even if \ $\G \neq \G^{\dagger}$ \ [non-hermiticity],
quantity $G(x)$
computed by means of eq.(3) will be always {\em real}.
 \ The only difference with respect to the
case of hermitian
operators is that the mean value $<G>$ and the eigenvalues $G_{i}$
will not be real; but this does {\em not} necessarily mean that $G$ is
unobservable. \ From the very definition of eigenstate in quantum mechanics,
in fact, an eigenstate of $\G$ from a ``local point of view'' is
characterized by a function $G(x)$ {\em
uniformly distributed (spatially homogeneous and constant in time)}:
 \ $G(x) = G_{i}$ \ for any $x$. \ \
Then, one can conclude$^{[1]}$ that necessary condition for the hermiticity
of $G$, and the consequent
existence of real eigenvalues $G_{i}$, is the possibility
of creating and observing a {\em uniform distribution for quantity} $G$ in
correspondence with the chosen eigenstates. \  The inverse does not
hold: it is possible
to have locally uniform quantities not corresponding to hermitian
operators. \ A noticeable example of this occurrence is given by the
non-hermitian {\em velocity operator} proposed below.
In spite of its non-hermiticity, we shall see for
{\em plane waves\/} [$\imp =$constant] that its non-hermitian part
will give no contribution, so that the velocity field will be
locally uniform and equal to $\imp/m$.\\

{\bf 2. -- A new non-relativistic velocity operator endowed with
zitterbewegung}\\

\h In the framework of the Pauli geometric algebra, the local velocity
is obtained
from the usual operator $\impo/m$, once it is ``translated" into the new algebraic
language. \ Thus we shall have, following the standard rules for that
translation,
$$
i\hbar \longrightarrow {\bf i}2{\wide {\sbf}} \equiv {\bf i}\hbar\sigbf\; ,
\eqno{(\rm 6a)}
$$
$$
\impo/m \equiv i\hbar\vecna/m \longrightarrow \vecna{\bf i}\hbar\sigbf/m
\eqno{(\rm 6b)}
$$
where ${\wide {\sbf}}$ represents the {\em spin vector operator},
$\sigbf$ indicates the usual $2 \times 2$ Pauli matrices,
and ${\bf i}$ the Pauli algebra {\em pseudoscalar unity} (which corresponds to
the matrix $\sig_x\sig_y\sig_z$, so that ${\bf i}^2 = - \1$). \
Therefore, we can write for the velocity field:
\setcounter{equation}{6}
\bb
\vbf (x) = -{1\over m}\,\frac{\langle\psi\vecna{\bf i}\hbar\sigbf\psip\rangle_0}{\psi\psip}\; .
\ee
The ``tensorial version'' of this expression is the velocity
operator:$^{\# 1}$
\footnotetext{$^{\# 1}$ Let us recall that, with regard to the
{\em vectorial} basis $\sigbf^1 , \sigbf^2 , \sigbf^3$
of the Pauli multivector algebra, we have by definition: \  $\vecna\equiv
\sigbf^i\nabla_i$, \ indicating by $\nabla_i$ the $i$-th component of vector
$\vecna$.}
\bb
\vo = {i\hbar\over m}\sigbf\,(\vecna\cdot\sigbf) \ ,
\ee

Due to the mathematical identity
\bb
\sigbf\,(\abf\cdot\sigbf) \equiv \abf + i\,\abf\times\sigbf\; ,
\ee
$\abf$ being a generic 3-vector, we shall finally get:
\bb
\vo = -\frac{i\hbar}{m}\,\vecna + \frac{\hbar}{m}\,(\vecna\times\sigbf)
\equiv \frac{\impo}{m} + \frac{i}{m}\,(\impo\times\sigbf)\ ,
\ee
where $\impo$ and $\sigbf$ commute, making ininfluent
the order in which they appear in the product.

\h The above operator results to be composed by a hermitian part,
$\impo/m$, and by a non-hermitian part, $i\,(\impo\times\sigbf)/m$.
 \ The hermitian part reduces to the ordinary (but ``incomplete'': see below)
non-relativistic
operator for wave-mechanics, usually written as \
$ i [\Ho\, , \,\xo] / \hbar \equiv
 i [{\impo}^2 / 2m\, , \,\xo] / \hbar$. \
The non-hermitian part is strictly related
to the so-called {\em zitterbewegung} (zbw),$^{[12,13]}$ which is the
spin motion, or ``internal'' motion ---since it is observed in the CMF,---
expected to exist for spinning
particles. Such a motion of an internal ``constituent" $\cal Q$ appears
only for particles endowed with spin,$^{[13]}$ and is to be added to the
drift--translational, or ``external'', motion of the CM, $\,\imp/m$ (which is
the only one occurring for scalar particles). \ In the Dirac theory, indeed,
the operators $\vo$ and $\impo$ are not parallel:
\bb
\vo\neq\impo/m\; .
\ee
Moreover, while \ $[\wide{\imp},
\wide{H}]=0$ \ so that $\imp$ is a conserved quantity,
$\vbf$ is {\em not} a constant of the motion: \ $[\wide{\vbf}, \wide{H}]\neq
0$ \ (quantity \ $\wide{\vbf} \equiv \albf \equiv \ga^0\gabf$ \ being
the usual
vector matrix of the Dirac theory). \ \ Let us notice that
in case of zbw it is highly convenient$^{[12,13]}$ to split the motion
variables as follows (the dot meaning derivation with respect to time):
\bb
\xo \equiv \xio + \Xo \; ; \ \ \ \xdo \equiv \vo = \wo + \Vo \; ,
\ee
where $\xio$ and $\wo \equiv \xido$ describe the motion of the CM,
whilst $\Xo$ and $\Vo \equiv \Xdo$ describe the zbw motion. \
From an electrodynamical point of view, the conserved electric current is
associated with the
trajectories of $\cal Q$ \ (i.e., with $\xo$), whilst the center of the
particle Coulomb field ---obtained$^{[14]}$ via a time average
over the field produced by the quickly oscillating charge---
coincides with the particle CM (i.e., with $\wo$) and therefore, for free
particles, with the geometrical center of the helical trajectory.

\h As a consequence, it is $\cal Q$ which performs the {\em total motion},
while the CM undergoes the {\em mean motion} only. \ The resulting electron
can be regarded as ``extended-like'',$^{[13]}$ because of the existence
in the CMF of an internal spin motion.

\h As required by eq.(11), one has to assume the existence of zbw also
in the standard Dirac theory. In fact, the above
decomposition for the total motion comes out in two well-known relativistic
quantum--mechanical procedures: namely, in the
{\em Gordon decomposition} of the Dirac current, and in the
{\em decomposition of the Dirac
velocity operator and Dirac position operator}
proposed by Schr\"odinger in his pioneering works.$^{[15]}$

\h The Gordon decomposition of the Dirac current reads (hereafter we shall
chose units such that $c=1$):
\bb
\psib\ga^\mu\psi = \frac{1}{2m}\,[\psib\po^\mu\psi - (\po^\mu\psib)\psi]
- \frac{i}{m}\po_\nu\,(\psib S^{\mu\nu}\psi)\; ,
\ee
$\psib$ being the ``adjoint'' spinor of $\psi$; \ quantity \ $\po^\mu
\equiv i\pa^\mu$ \ the 4-dimensional impulse operator; \ and $S^{\mu\nu}
\equiv \frac{i}{4}\,(\ga^\mu\ga^\nu - \ga^\nu\ga^\mu)$
the spin-tensor operator. \
The ordinary interpretation of eq.(13) is in total analogy with the
decomposition given in eq.(10).
The first term in the r.h.s. of eq.(13) results to be associated with
the translational motion of the CM (the {\em scalar} part of the current,
corresponding to the traditional Klein--Gordon current).  By contrast, the
second term in the r.h.s. results related to the existence
of spin, and describes the zbw motion.

\h In the above quoted papers, Schr\"odinger started from the Heisenberg
equation for the time evolution of the acceleration operator in Dirac theory
\bb
\ao \equiv \frac{\drm\vo}{\drm t} \ug {i\over\hbar}\,[\Ho, \vo] \ug
{i\over\hbar}\,2(\Ho\vo - \impo)\; ;
\ee
where $\Ho$ is equal as usual to \ $\vo$$\cdot$$\impo + \bt\,m$
 \ (where $\vo\equiv\albf$). \
By integrating once this operator equation over time, he obtained:
\bb
\vo \ug \Ho^{-1}\impo + \eto (0)\,\erm^{-2i\Ho t/\hbar}
\qquad  (\eto \equiv \vo - \Ho^{-1}\impo)\; .
\ee
After a little algebra, we may get a more interesting form for the velocity
decomposition:
\bb
\vo \ug \Ho^{-1}\impo - {1\over 2}i\hbar\,\Ho^{-1}\ao\; .
\ee
By integrating a second time, Schr\"odinger ended up also with the spatial
part of the decomposition:
\bb
\xo \equiv \xio + \Xo \ ,
\ee
where we have
\bb
\xio = \ro + \Ho^{-1}\impo t \ ,
\ee
linked to the motion of the CM, \ and
\bb
\Xo = {1\over 2}i\hbar\,\eto \Ho^{-1} \ ,
\ee
linked to the zbw motion.

\h We can therefore consider decomposition (10) of our velocity operator
as the
{\em non-relativistic analogue
of decomposition} (16) {\em of the relativistic velocity operator}. \ It is
not at all surprising that (besides spin and the related intrinsic magnetic
moment) also another ``spin effect", zbw,
does not vanish in the
non-relativistic limit, i.e., for small velocities of the CM \
[$\imp\longrightarrow 0$]. \ \ Therefore also the Schr\"odinger electron,
being endowed with a zbw motion, does actually show its spinning
nature, and is not a ``scalar'' particle (as often assumed)! \ As a
matter of fact,
when constructing atoms, we have necessarily to introduce ``by hand" the
Pauli exclusion
principle; and in the Schr\"odinger equation the Planck constant $\hbar$
implicitly denounces the presence of spin. \ \
In ref.[16] we have proved that the non-hermitian (zbw) part of
our velocity field gives origin to the {\em quantum potential} of the
Madelung fluid, as well as to the related
{\em zero--point energy} of the Schr\"odinger theory.\\

{\bf 3. -- The velocity field of the Madelung fluid}\\

\h Even if the above operator $\vo$ is not hermitian, the local velocity
$\vbf (x)$, as said above, will result to be always a real quantity. Let us
see it by inserting eq.(10) into definition (3).

\h A spinning non-relativistic particle can be represented by means of a
Pauli 2-components spinor:
\bb
\psi \equiv \sqrt{\rho}\> \Phi\, ,
\ee
where, if we require $|\psi|^2 = \rho$, quantity $\Phi$ must obey the
normalization constraint
\bb
\Phi^{\age}\Phi = 1 \ .
\ee
Inserting the factorization (20) into definition (3), we have:
\bb
\sbf \equiv \frac{{\rm Re}\,\{\psip\hh\sigbf\psi\}}{\psip\psi} \equiv {\hbar/2}
\Phi^{\age}\sigbf\Phi\; ,
\ee
if $G$ is the spin vector; \ and
\bb
\imp \equiv \frac{{\rm Re}\,\{\psip(-i\hbar\vecna)\psi\}}{\psip\psi} =
\frac{i\hbar}{2m} [(\vecna \Phi^{\age})\Phi - \Phi^{\age} \vecna \Phi]\; ,
\ee
if $G$ is the impulse.

\h In the most general case ({\em Pauli} generalization of Schr\"odinger
theory), when it is present an external potential $\Abf\neq 0$,
we have to replace in the translational term
of expression (10), as a ``minimal prescription",
the canonic impulse $\impo$ by the kinetic impulse $\impo - e\Abf$; \ where
$e$ is the particle electric charge. \  Let us substitute in eq.(3)
this ``generalized'' velocity operator for $G$, and factorization (20)
for $\psi$: \ we finally get the following decomposition for the velocity
field of the non-relativistic quantum fluid:
\bb
\vbf = \frac{\imp - e\Abf}{m} + \frac{\vecna\times(\rho\sbf)}{m\rho}\; .
\ee
This expression may be considered as the non-relativistic analogue
of the Gordon decomposition (13) of the Dirac current.

\h We want to stress that decomposition (24), just now derived by means of
definition (3), may be also obtained within
standard wave-mechanics: and this will be a further test of the validiy
of our operator. It is sufficient, in fact, to take the familiar expression
of the Pauli current (i.e., the non-relativistic limit of the Gordon
decomposition$^{[17]}$)
\bb
\jbf \equiv \rho\vbf = \frac{i\hbar}{2m}[(\vecna \psi^{\age})\psi - \psi^{\age} \vecna \psi] -
\frac{e\Abf}{m}\psi^{\age}\psi +
\frac{\hbar}{2m}\vecna \times (\psi^{\age} \sigbf\psi)\; ,
\ee
and to insert into it factorization (20) in place of $\psi$,  for
obtaining the velocity distribution given by eq.(24).\\

\h Let us single out in the total velocity field the zbw--component:

\bb
\Vbf \equiv \frac{\vecna\times (\rho\sbf)}{m\rho}\; .
\ee
It is furtherly decomposable in two distinct parts:

\h A) \ \ \ $\Vbf_1 \equiv (\vecna\rho\times\sbf)/m\rho$\\
due to the presence of the gradient of $\rho$, this term refers to local
motions, in which the constant density surfaces [$\vecna\rho = 0$] do
rotate around
the spin axis; and it vanishes identically for the plane waves [$\imp$ =
constant], for which $\vecna\rho = 0$;

\h B) \ \ \  $\Vbf_2 \equiv ({\rm rot}\sbf)/m$\\
such a term does not depend on the density $\rho$ and is different from 0
only in presence of external magnetic fields (the spin vector precedes,
and the wave-functions are {\em not} spin eigenstates).

\h The {\em Schr\"odinger theory} is of course a particular case in
the present Pauli framework, corresponding to $\Abf =0$ (and then to zero
magnetic field) and to a {\em uniform local spin vector} with no precession.
The wave-function is then a {\em spin eigenstate} and may be factorized
as the product of a ``scalar" part
$\sqrt{\rho}e^{i\varphi}$ and of a ``spin" part $\chi$ (a 2-components spinor):
\bb
\psi \equiv \sqrt{\rho}\, e^{i\frac{\varphi}{\hbar}}\chi\; ,
\ee
quantity $\chi$ being {\em constant} in space and time. \ Let us underline
that, even if $\sbf = \chi^{\age}\hh\sigbf\chi =$ constant, {\em in the
Schr\"odinger case zbw does not vanish} ---except for the
unrealistic case of plane waves---  while the velocity $\Vbf$ reduces to
\bb
\Vbf = \Vbf_1 \equiv \frac{(\vecna\rho\times\sbf)}{m\rho}\; .
\ee
All this does actually contribute to remove some difficulties remaining in
the classical
representation of the particle motion, and in the interpretation of the
particle energies, for some stationary solutions of the Schr\"odinger
equation. For instance, let us refer ourselves to the stationary
states of a particle inside a box or of the harmonic oscillator, or the
$l=0$ eigenstates of the hydrogen atom. In all these cases the wave-function
results to be {\em real}  ($\varphi$ is uniform and equal to a constant which
for ``global gauge invariance" may be assumed equal to zero), and
therefore {\em the velocity obtained from standard quantum mechanics
$\imp/m \equiv \vecna\varphi/m$ is zero} everywhere, at any time. \ As it
was first remarked by Einstein and Perrin and by de Broglie,$^{[18]}$ such a
result {\em seems to be really in contrast} from a classical point of view
{\em with the non-vanishing of the energy eigenvalue for those stationary
states}. \ But we now know, from our previous analysis,
that $\imp/m$ is the {\em
mean} velocity, describing only the motion of the CM, whilst the
zbw--component $\Vbf$ ---which depends on the $\rho$--gradient, and not on the
phase--gradient of $\psi$--- does not identically vanish, thus implying
an internal motion around the spin axis.\\

\h Let us finally analyse the velocity distribution (24) for $\Abf = 0$,
and stress some interesting properties of its.

\h Since the rotor of a gradient is identically zero, we shall have
${\rm rot}\, \imp = 0$; and since $\sbf$ is furthermore uniform, it will
also be $\vecna\sbf = 0$. \ As a consequence, by
employing the known property of the double vectorial product
\bb
\abf\times(\bbf\times\cbf)\; = \; (\abf\cdot\cbf)\,\bbf - (\abf\cdot\bbf)
\, \cbf \; ,
\ee
we can show that the local rotational properties of the Madelung fluid are
actually given by the following expression:
\bb
{\rm rot} \: \vbf \; = \; \frac{1}{m}\,\left[\left(\frac{{\vecna}\rho}{\rho}
\right)^2
- \frac{\triangle \rho}{\rho}\right]\,\sbf \; .
\ee
Moreover, the zbw current field $\rho\Vbf$ results to be {\em solenoidal} (and this
happens
also in the most general case of the Pauli fluid with a non-uniform $\sbf$):
\bb
{\rm div}\,(\rho\Vbf) \equiv {\rm div}\,[{\rm rot}(\rho\sbf)]/m \; = \; 0\; .
\ee
The flux stream-lines will be {\em closed lines}, as in the magnetic field
case: therefore, we expect {\em the zbw motion to be limited, finite,
periodical}.
The explicit calculations performed by us for the known
Barut--Zanghi model$^{[13]}$ did really lead in the CMF of the electron to
closed periodical motions. \
The total motion of the system will be helical around the
$\imp$-direction. We can see eventually that the {\em local angular
velocity $\ombf$ is parallel to the spin vector $\sbf$}:
\bb
\ombf = {1\over 2}\,{\rm rot} \, \vbf \; = \; \frac{1}{2m}\,\left[
\left(\frac{{\vecna}\rho}{\rho}\right)^2
- \frac{\triangle \rho}{\rho}\right]\,\sbf \; .
\ee

\vs{2.0 cm}

\newpage

{\bf Acknowledgements}\\

This work is dedicated to the memory of Asim O. Barut. \  The authors wish to
acknowledge stimulating discussions with J. Keller, F. Raciti, W.A. Rodrigues
and J. Vaz.
\ For the kind cooperation, thanks are also due to A. Agresti, G. Andronico,
I. Arag\'on, M. Baldo, A. Bonasera, M. Borrometi, A. Bugini, F. Catara,
L. D'Amico, G. Dimartino, C. Dipietro, M. Di Toro, G. Marchesini, L. Lo Monaco,
R.L. Monaco, Z. Oziewicz, M. Pignanelli, G.M. Prosperi,
M. Sambataro, S. Sambataro, M. Scivoletto, R. Sgarlata, R. Turrisi and
M.T. Vasconselos.

\end{document}